# Detection and Feeder Identification of the High Impedance Fault at Distribution Networks Based on Synchronous Waveform Distortions

Mingjie Wei, Fang Shi, Hengxu Zhang, Weijiang Chen, Bingyin Xu

*Abstract*—Diagnosis of high impedance fault (HIF) is a challenge for nowadays distribution network protections. The fault current of a HIF is much lower than that of a normal load, and fault feature is significantly affected by fault scenarios. A detection and feeder identification algorithm for HIFs is proposed in this paper, based on the high-resolution and synchronous waveform data. In the algorithm, an interval slope is defined to describe the waveform distortions, which guarantees a uniform feature description under various HIF nonlinearities and noise interferences. For three typical types of network neutrals, i.e., isolated neutral, resonant neutral, and low-resistor-earthed neutral, differences of the distorted components between the zero-sequence currents of healthy and faulty feeders are mathematically deduced, respectively. As a result, the proposed criterion, which is based on the distortion relationships between zero-sequence currents of feeders and the zero-sequence voltage at the substation, is theoretically supported. 28 HIFs grounded to various materials are tested in a 10kV distribution network with three neutral types, and are utilized to verify the effectiveness of the proposed algorithm.

## I. INTRODUCTION

HIGH impedance faults (HIFs) account for more than 10% of the total fault number at medium-voltage (MV) distribution networks [1]. The amplitudes of HIF currents are generally lower than 50A [2], which is much lower than normal load currents. According to the early staged tests by Texas A&M University, the detecting rate of HIFs by conventional overcurrent relays is only about 17.5% [3]. Although these low-current faults do not damage power system components, their long-time existence could raise the risks of fire and shock accidents. The severe fire hazards recently occurring in Australia, the United States, and Brazil, are confirmed to initially result from HIFs [4], which emphasizes the necessity and urgency of HIF diagnosis. In general, HIFs are the single-line-to-ground (SLG) faults happening in overhead lines and caused by the following accidents:

1) a line is broken and falls to the ground that is with high impedance surface material, like soil, concrete, asphalt, and grass.

2) an intact line sags to the ground and touches the above high impedance surface material.

3) an intact line is touched by an object that is with high impedance material, such as a tree limb.

When the conductor of a faulted line establishes electrical conduction with ground medium, arc always ignites [5] as many air gaps exist between the conductor and ground, or inside the uncompacted ground materials. As a result, the nonlinearity generated by the arcing process is a prominent characteristic of HIF. If conductors touch some materials with extremely high impedance, fault currents can be restricted below 1 A. It further increases difficulties in guaranteeing the detection reliability under interferences from load currents and background noises. Algorithms are also strictly required to be able to distinguish HIFs from non-fault conditions, so as to avoid unnecessary outages. Besides, the diverse characteristics of HIFs affected by materials, humidity, and system neutrals [6], etc., also bring trouble.

For distribution networks without widespread installations of advanced meters, faulty feeder identification at substations is of paramount importance to realize fault tracking and isolation. The feeder identification is commonly triggered by fault detection.

The detection of HIFs has been researched for about 40 years, and algorithms are generally classified into two categories: the rule-based approaches, and pattern-recognition-based approaches. On the one hand, the rule-based approaches firstly establish the equivalent circuits of networks, and then mathematically deduce the amplitude or phase changes of voltage, current, impedance, or admittance. These methods are theoretically supported and logically demonstrated, but usually only applicable to the networks with specific neutrals, like the isolated neutral [7], [8], compensated (resonant) neutral [8], and solidly-earthed neutral [9]. In addition, rule-based approaches often neglect the measuring errors caused by the nonlinearities of HIFs, which may result in malfunctions. For example, in a stable HIF, the phase calculated by Fourier transform generally lags behind its real values.

On the other hand, the pattern-recognition-based approaches detect HIFs by taking advantage of arc nonlinearities, including the anomalies of harmonics and the distortions of waveform shapes. The harmonic based algorithms include those utilizing even harmonics [10], odd harmonics [11], inter-harmonics [12], and high-frequency harmonics [13], whereas the waveform shape based algorithms include those utilizing the asymmetry between half-cycles [14] and the waveform distortions near the

This work was supported by the National Key R&D Program of China (2017YFB0902800) and the Science and Technology Project of State Grid Corporation of China (52094017003D).

Mingjie Wei, Fang Shi and Hengxu Zhang are all with the Key Laboratory of Power System Intelligent Dispatch and Control Ministry of Education, Shandong University, Jinan, 250061, China (e-mail: zhanghx@sdu.edu.cn).

Weijiang Chen is with the State Grid Corporation of China (e-mail:weijiang-chen@sgcc.cn).

Bingyin Xu is with the School of Electrical and Electric Engineering, Shandong University of Technology, Zibo, 255000, China (xuby@vip.163.com).

zero-crossings of currents [15]-[16]. With the development of signal analysis technology, tools utilized in HIF detection have been continually updated. For example, finite impulse response filter [12], mathematical morphology [16], Kalman filtering [17], wavelet transform [18], and the improved methods for adaptive time-frequency analysis [19] are used for feature extraction. Meanwhile, intelligent algorithms like support vector machine [1], decision tree [17], fuzzy theorem [20], expert system [21], and artificial neural network [22] are used to act as classifiers to distinguish HIFs from non-fault conditions.

However, limitations still exist for the above efforts:

1) for intelligent algorithms, representative real-world training data is lacking due to the difficulty in recording HIFs in reality. It makes the dependability of these algorithms questionable.

2) for low-order harmonic based algorithms, the detection sensitivity of HIFs with fault current less than several amperes cannot be guaranteed due to interference from large load currents and intense noises.

3) for high-order harmonic based algorithms, the high-frequency components of HIFs cannot propagate well when shunt capacitors are under operations [22] or when the system neutral is earthed with a resistor [23]. Meanwhile, high-frequency harmonics are much weaker if conductors touch bad insulation materials under wet conditions [24], and in this case, harmonics can be greatly interfered with by noises. Besides, it is also challenging to distinguish high-frequency components of HIFs from that of some non-fault events, like nonlinear loads and inrush currents.

For the feeder identification, related approaches were initially researched for low impedance faults (LIFs) at networks with non-solidly earthed neutral, where the fault currents are also limited. These techniques are mature and have been industrially applied. For example, a method based on transient components in a selected frequency band (SFB) [25] has been applied over 15 years in China. However, for HIFs, the transient high-frequency components are significantly suppressed by the large grounding impedances [26]. In [7] and [26], two rule-based approaches are proposed to identify HIF feeders respectively with zero-sequence capacitances and phases of zero-sequence currents. Two pattern-recognition-based algorithms in [15] and [23] identify HIF feeders with waveform distortions. However, their sensitivities are all restricted by neutral types and distortion diversity.

Synchronous measuring is the basis to guarantee the feasibility of the above HIF feeder identification approaches. In our previous work [18], [27], high-resolution and synchronous data monitored by advanced fault recorders (developed since 2011) are utilized. In this paper, an algorithm is proposed to detect HIFs and identify the faulty feeders with synchronous waveforms. With field HIF data achieved in a 10kV distribution network, the distortion diversity caused by different fault scenarios and noise interferences is firstly analyzed. By deducing the differences of waveform distortions between feeders, an integrated algorithm based on a definition of interval slope is proposed to detect HIFs and identify the faulty feeders. The algorithm is demonstrated to be useful for three common neutral networks, and reliable under various distortion shapes and noisy environments.

The rest of this paper is organized as follows: In Section II, characteristics of HIF nonlinearity are analyzed. Differences of nonlinear distortions between feeders are deduced in Section III, based on which a feeder identification algorithm is proposed in Section IV. In Section V, effectiveness of the proposed algorithm is verified. Finally, conclusions are drawn in Section VI.

## II. CHARACTERISTIC OF HIF NONLINEARITY

A certain number of HIFs are artificially experimented in a 10kV 50Hz distribution network (Fig.1), and used for analysis in this paper. There are 4 feeders under operations. Herein, L1 is an overhead line, L3, L4 are the underground cables, and L2 is a hybrid line with the overhead part (0.5 km) between M2 and M5. The fault position is about 80 meters in electrical distance from M2. HIFs are tested by earthing the conductors to different materials and under different humidity. Three neutrals shown in Fig.1 are all tested. Signals are measured by the synchronous digital fault recorders [18], [27] deployed at the beginning of each feeder with sampling frequencies of 6.4 kHz.

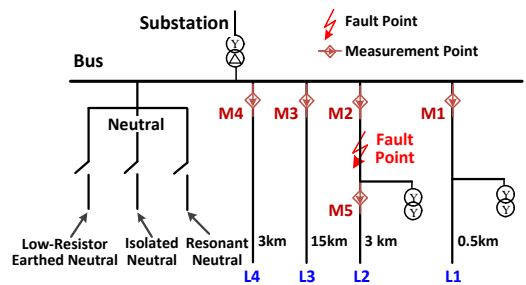

Fig.1 Topology of a 10 kV distribution network.

Arcing intermittence is a common phenomenon caused by the interruption of arc due to disturbances. It could result from the bounce of conductor, the blow of wind, the expulsion and refill of moisture, the charring of tree limb, the melt of ground material, the generation of smoke or steam, and the movement of material particles or pieces, etc. However, the intermittence does not happen all the time, and the arcing process could hold stable for minutes at the beginning of HIFs. It indicates that only the randomness algorithm [2] cannot always guarantee a fast fault detection.

For a stable HIF, arc nonlinearity during the zero-crossings of current is a prominent feature. The nonlinearity is caused not only by the ionization in air but also by the plasma propagation through the inside dielectric of ground [15], [24]. Under the circumstance, the material, humidity, and internal structure of the ground can all affect the physical thermal conversions of arc, and further affect the arc diameter and the arc resistance. As a result, currents of HIFs present a diversity of waveform distortions (Fig.2) under different conditions. The evaluation of an algorithm thereby needs to consider the cases when:

1) the high-frequency harmonics are weak for some smooth or slight waveform distortions like Fig.2 (d) and (e);

2) the distortion 'offsets' are various (see the relative positions of the midpoint of distortion interval in Fig.2) due to different capabilities of energy dissipation;

3) the waveforms exhibit ineffective distortions, like Fig.2

(g) and (h), where the impulse signals generated by arcing intermittence may disturb the description of waveform shape.

● Midpoint of distortion interval ┈ Zero-crossing ◯ Ineffective distortions

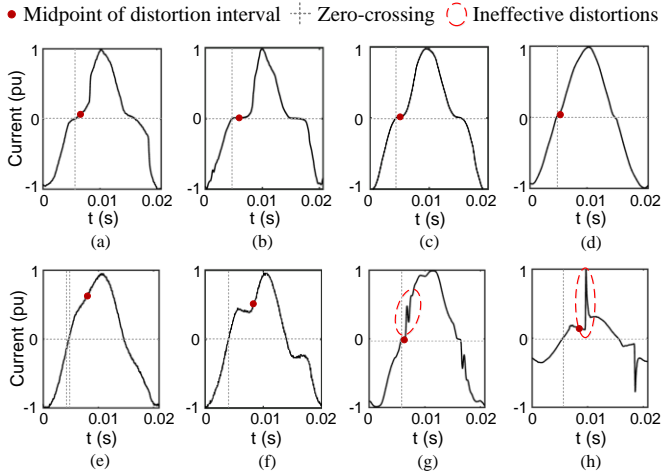

Fig.2 Current waveforms of the HIFs tested in a 10 kV distribution network: (a) wet asphalt concrete, isolated neutral; (b) wet soil, low-resistor-earthed neutral; (c) dry soil, isolated neutral; (d) wet cement, low-resistor-earthed neutral; (e) dry grass, resonant neutral; (f) wet grass, resonant neutral; (g) dry cement pole, isolated neutral; (h) dry soil, resonant neutral.

## III. DISTORTION DIFFERENCES BETWEEN FEEDERS

In this section, distortion differences of healthy and faulty feeders are theoretically analyzed for networks with different neutrals. Zero-sequence current is used for analysis as it is small in the pre-fault state and immune to load side behaviors [7], [15].

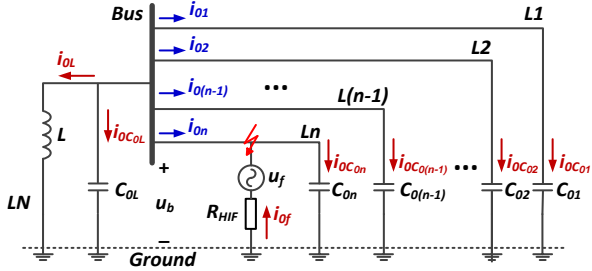

Fig.3 Equivalent zero-sequence circuit of a multi-feeder resonant network.

### A. Resonant Neutral

Fig.3 is an equivalent zero-sequence circuit of a multi-feeder distribution resonant network [26], where

$C_{0i}, C_{0L}$   Equivalent zero-sequence grounding capacitance of supply feeder $i$ ($i=1,2,..,n$) and the transformer feeder

$L$   Equivalent zero-sequence inductance of the Petersen coil (3 times the coil inductance)

$R_{HIF}$   Equivalent zero-sequence resistance of fault (3 times the grounding resistance)

$u_f$   Equivalent virtual voltage source, which has the same magnitude with and the opposite phase to the pre-fault phase-to-ground voltage at the fault point

$u_{0b}$   Zero-sequence voltage at the substation bus

$i_{0f}$   Zero-sequence fault current

$i_{0i}$   Zero-sequence current of feeder $i$

$i_{0C_{oi}}, i_{0C_{oL}}$   Zero-sequence grounding capacitance current of feeders

$i_{0L}$   Zero-sequence current flowing through the Petersen coil

Suppose that a HIF happens in feeder $n$, and the zero-sequence current in the faulty feeder $i_{0n}$ is expressed as:

$$i_{0n} = -i_{0f} + i_{0C_{on}} = -\left(i_{0L} + i_{0C_{oL}} + \sum_{i=1}^{n} i_{0C_{oi}}\right) + i_{0C_{on}} \quad (1)$$

In the following, $i_{0C_{oL}}$ is neglected as it is commonly too small compared to $i_{0L}$.

For a stable HIF, distortion is caused by the nonlinear part of fault resistance and happens periodically at power frequency. Therefore, the fault current $i_{0f}$ can be divided into a distorted component $\Delta i_{0f,dist}^{(\varphi-\pi)}$ and a sinusoidal component $i_{0f,sinu}^{(\varphi-\pi)}$. The $\varphi-\pi$ represents the phase of $i_{0f}$ and the above two components, where $\varphi$ is the phase of $i_{0C_{oi}}$.

It is the same for $i_{0L}$, $i_{0C_{oi}}$ and $i_{0n}$, where their distorted and sinusoidal components can be represented by $\Delta i_{0L,dist}^{(\varphi-\pi)}$, $\Delta i_{C_{oi},dist}^{(\varphi)}$, $\Delta i_{0n,dist}^{(\varphi-\pi)}$, and $i_{0L,sinu}^{(\varphi-\pi)}$, $i_{0C_{oi},sinu}^{(\varphi)}$, $i_{0n,sinu}^{(\varphi)}$, respectively. As a result, (1) can be expanded as:

$$\begin{aligned} i_{0n} &= -\left(i_{0L} + \sum_{i=1}^{n} i_{0C_{oi}}\right) + i_{0C_{on}} = -\left(i_{0L} + \sum_{i=1}^{n-1} i_{0C_{oi}}\right) \\ &= -\left[\left(i_{0L,sinu}^{(\varphi-\pi)} + \Delta i_{0L,dist}^{(\varphi-\pi)}\right) + \sum_{i=1}^{n-1}\left(i_{0C_{oi},sinu}^{(\varphi)} + \Delta i_{0C_{oi},dist}^{(\varphi)}\right)\right] \\ &= \left(I_{ML} - \sum_{i=1}^{n-1} I_{MC_{0i}}\right)\sin(\omega t + \varphi) + \left(-\Delta i_{0L,dist}^{(\varphi-\pi)} - \sum_{i=1}^{n-1}\Delta i_{0C_{oi},dist}^{(\varphi)}\right) \\ &= i_{0n,sinu}^{(\varphi)} + \Delta i_{0n,dist}^{(\varphi)} \end{aligned} \quad (2)$$

where, $I_{ML} > \sum_{i=1}^{n} I_{MC_{0i}} > \sum_{i=1}^{n-1} I_{MC_{0i}}$. $I_{MC_{0i}}$ and $I_{ML}$ are the peak values of $i_{0C_{oi},sinu}^{(\varphi)}$ and $i_{0L,sinu}^{(\varphi-\pi)}$. $\omega$ represents the radian power frequency and equals $100\pi$ rad/s in this paper.

According to Fig.3, zero-sequence voltage $u_{0b}$ is also expressed in the form of distorted and sinusoidal components:

$$u_{0b} = u_{0b,sinu}^{(\varphi-\pi/2)} + \Delta u_{0b,dist}^{(\varphi-\pi/2)} = \frac{1}{C_{0i}}\int i_{0C_{oi}} dt = L\frac{di_{0L}}{dt} \quad (3)$$

$$\begin{aligned} \Delta u_{0b,dist}^{(\varphi-\pi/2)} &= \frac{1}{C_{01}}\int \Delta i_{0C_{01},dist}^{(\varphi)} dt = \cdots = \frac{1}{C_{0n}}\int \Delta i_{0C_{on},dist}^{(\varphi)} dt \\ &= \frac{1}{C_{0\Sigma}}\int \sum_{i=1}^{n}\Delta i_{0C_{oi},dist}^{(\varphi)} dt = L\frac{d\Delta i_{0L,dist}^{(\varphi-\pi)}}{dt} \end{aligned} \quad (4)$$

where $C_{0\Sigma} = \sum_{i=1}^{n} C_{0i}$; $u_{0b,sinu}^{(\varphi-\pi/2)}$ and $\Delta u_{0b,dist}^{(\varphi-\pi/2)}$ are the sinusoidal and distorted components of $u_{0b}$, which are both with the phase of $\varphi-\pi/2$.

Then, according to (1) and (4), $\Delta i_{0f,dist}^{(\varphi-\pi)}$ is expressed as:

$$\begin{aligned} \Delta i_{0f,dist}^{(\varphi-\pi)} &= \Delta i_{0L,dist}^{(\varphi-\pi)} + \sum_{i=1}^{n}\Delta i_{0C_{oi},dist}^{(\varphi)} \\ &= \Delta i_{0L,dist}^{(\varphi-\pi)} + LC_{0\Sigma}\frac{d^2\Delta i_{0L,dist}^{(\varphi-\pi)}}{dt^2} \end{aligned} \quad (5)$$

Fig.4 shows a $i_{0f}$ in the range of $\omega t \in [-\varphi-\pi, -\varphi+\pi)$, as well as its sinusoidal and distorted components. To figure out the inherent differences between the distortions in healthy and faulty feeders, $\Delta i_{0L,dist}^{(\varphi-\pi)}$, $\Delta i_{0C_{oi},dist}^{(\varphi)}$, and $\Delta i_{0n,dist}^{(\varphi)}$ need to be theoretically analyzed, which is impractical when their ex-pressions are unknown. For ease of interpretation, $\Delta i_{0f,dist}^{(\varphi-\pi)}$ is simplified as a piecewise function $f(\omega t)_{0f}^{(\varphi-\pi)}$, which is also presented in Fig.4. Specifically, the $f(\omega t)_{0f}^{(\varphi-\pi)}$ in the range of $\omega t \in [-\varphi-\pi, -\varphi-\frac{\pi}{2})$ is expressed as:

$$f(\omega t)_{0f}^{(\varphi-\pi)} = -e^{\frac{\tau}{\omega}(\omega t+k)} \cdot I_{fM,dist} \sin[2(\omega t+\varphi-\pi)] \quad (6)$$

where, $-1 < \tau < 0$ and $k = \varphi + \pi$; $I_{fM,dist}$ is set to make the peak value of $f(\omega t)_{0f}^{(\varphi-\pi)}$ be equal to that of $\Delta i_{0f,dist}^{(\varphi-\pi)}$.

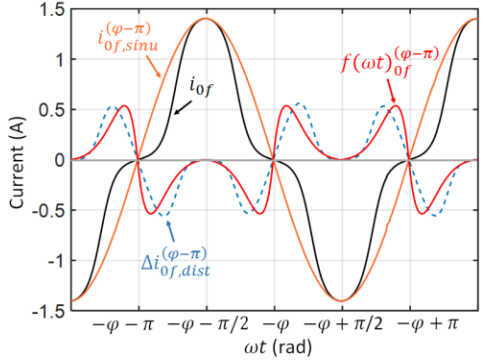

Fig.4 Relationship between the distorted and sinusoidal components of $i_{0f}$.

According to Fig.4, if neglecting the distortion offset, $f(\omega t)_{0f}^{(\varphi-\pi)}$ is supposed to be axial-symmetry to $\omega t = -\varphi - \frac{\pi}{2}$, and the following two equations are satisfied:

$$f(\omega t)_{0f}^{(\varphi-\pi)} = f\left[\left|\omega t - \left(-\varphi - \frac{\pi}{2}\right)\right|\right]_{0f}^{(\varphi-\pi)} \quad (7)$$

$$f(\omega t+\varphi)_{0f}^{(\varphi-\pi)} = -f(\omega t+\varphi+\pi)_{0f}^{(\varphi-\pi)}$$
$$\Rightarrow f(\omega t)_{0f}^{(\varphi-\pi)} = -f(\omega t)_{0f}^{(\varphi)} \quad (8)$$

According to (6), the equation in (5) can be written as a second-order non-homogeneous linear (NHL) equation:

$$\Delta i_{0L,dist}^{(\varphi-\pi)} + LC_{0\Sigma}\frac{d^2\Delta i_{0L,dist}^{(\varphi-\pi)}}{dt^2} = f(\omega t)_{0f}^{(\varphi-\pi)} \quad (9)$$

Take the interval of $\omega t \in [-\varphi-\pi, -\varphi-\frac{\pi}{2})$ as an example. Substitute (6) into (9) and the NHL equation is solved as:

$$\Delta i_{0L,dist}^{(\varphi-\pi)} \approx a_0 \cos\left(pt+\frac{p}{\omega}\varphi\right) + a_1 \sin\left(pt+\frac{p}{\omega}\varphi\right) + e^{\frac{\tau}{\omega}(\omega t+k)} \cdot A_L \sin[2(\omega t+\varphi-\pi)]$$

$$\Delta i_{0C_{0i},dist}^{(\varphi)} \approx -a_0 \cos\left(pt+\frac{p}{\omega}\varphi\right) - a_1 \sin\left(pt+\frac{p}{\omega}\varphi\right) + e^{\frac{\tau}{\omega}(\omega t+k)} \cdot A_{C_{0i}} \sin[2(\omega t+\varphi)] \quad (10)$$

$$p = \sqrt{\frac{1}{LC_{0\Sigma}}}, A_L = \frac{-I_{fM,dist}}{1-4\omega^2LC_{0\Sigma}}, A_{C_{0i}} = \frac{I_{fM,dist} \cdot 4\omega^2LC_{0i}}{1-4\omega^2LC_{0\Sigma}}$$

Similarly, solve the NHL equations in the other three intervals of $[-\varphi-\pi, -\varphi+\pi)$. In order to keep the continuity at the boundary of $-\varphi-\pi$ and $-\varphi$, coefficients $a_0$ and $a_1$ in (10) are both supposed to be 0. Finally, the zero-sequence currents of different feeders are expressed as:

$$i_{0L} = i_{0L,sinu}^{(\varphi-\pi)} + \Delta i_{0L,dist}^{(\varphi-\pi)} \quad (11)$$
$$\Delta i_{0L,dist}^{(\varphi-\pi)} = \frac{-A_L}{I_{fM,dist}}\Delta i_{0f,dist}^{(\varphi-\pi)}$$

$$i_{0i\,(i\neq n)} = i_{0C_{0i}} = i_{0C_{0i},sinu}^{(\varphi)} + \Delta i_{0C_{0i},dist}^{(\varphi)} \quad (12)$$
$$\Delta i_{0C_{0i},dist}^{(\varphi)} = \frac{-A_{C_{0i}}}{I_{fM,dist}}\Delta i_{0f,dist}^{(\varphi-\pi)} = \frac{A_{C_{0i}}}{I_{fM,dist}}\Delta i_{0f,dist}^{(\varphi)}$$

$$i_{0n} = i_{0n,sinu}^{(\varphi)} + \Delta i_{0n,dist}^{(\varphi)} \quad (13)$$
$$\Delta i_{0n,dist}^{(\varphi)} = -\left(\Delta i_{0L,dist}^{(\varphi-\pi)} + \sum_{i=1}^{n-1}\Delta i_{0C_{0i},dist}^{(\varphi)}\right)$$
$$= \frac{A_L}{I_{fM,dist}}\Delta i_{0f,dist}^{(\varphi-\pi)} + \sum_{i=1}^{n-1}\frac{A_{C_{0i}}}{I_{fM,dist}}\Delta i_{0f,dist}^{(\varphi-\pi)}$$
$$= \frac{1-4\omega^2L(C_{0\Sigma}-C_{0n})}{1-4\omega^2LC_{0\Sigma}}\Delta i_{0f,dist}^{(\varphi)}$$

where, $\Delta i_{0f,dist}^{(\varphi)}$ leads $\pi$ radians ahead of $\Delta i_{0f,dist}^{(\varphi-\pi)}$, and $\Delta i_{0f,dist}^{(\varphi)} = -\Delta i_{0f,dist}^{(\varphi-\pi)}$ according to (8). In a system with resonant neutral, a detuning index denoted as $v = 1 - 1/\omega^2LC_{0\Sigma}$ is used to describe the compensation level of the Petersen coil [26]. $v$ is generally in the range of $[-0.1, 0]$, so $\omega^2LC_{0\Sigma}$ is within $[0.9535, 1]$ and $4\omega^2LC_{0\Sigma}$ is larger than 1. Then, in (10), $A_L > 0$ and $A_{C_{0i}} < 0$.

Due to the nonlinear increase of arc resistance, the amplitudes of fault current $i_{0f}$ are always lower than its sinusoidal component $i_{0f,sinu}^{(\varphi-\pi)}$. That means the amplitudes of $i_{0f,sinu}^{(\varphi-\pi)}$ will decrease after being added by the distorted component $\Delta i_{0f,dist}^{(\varphi-\pi)}$. Under the circumstance, we say that $i_{0f}$ is obtained by the 'negative superposition' of $i_{0f,sinu}^{(\varphi-\pi)}$ and $\Delta i_{0f,dist}^{(\varphi-\pi)}$, otherwise, by the 'positive superposition'. With this knowledge, the distorted component in each feeder is analyzed as follows:

1) For the transformer feeder, according to (11), $\Delta i_{0L,dist}^{(\varphi-\pi)}$ has the opposite sign to $\Delta i_{0f,dist}^{(\varphi-\pi)}$. That means the effects on their respective sinusoidal components are also the opposite. Therefore, the amplitudes of $i_{0L,sinu}^{(\varphi-\pi)}$ increase after being added by $\Delta i_{0L,dist}^{(\varphi-\pi)}$, i.e., the positive superposition is claimed.

2) For the $i^{th}$ ($i \neq n$) healthy feeder, according to (12), $\Delta i_{0C_{0i},dist}^{(\varphi)}$ has the same sign with $\Delta i_{0f,dist}^{(\varphi-\pi)}$. As $i_{0C_{0i}}$ is with phase of $\varphi$, $\Delta i_{0f,dist}^{(\varphi-\pi)}$ is transformed to $\Delta i_{0f,dist}^{(\varphi)}$, which is with opposite sign to $\Delta i_{0C_{0i},dist}^{(\varphi)}$. Therefore, $i_{0i\,(i\neq n)}$ is obtained by positive superposition of $i_{0C_{0i},sinu}^{(\varphi)}$ and $\Delta i_{0C_{0i},dist}^{(\varphi)}$.

3) For the faulty ($n^{th}$) feeder, when the $1 - 4\omega^2L(C_{0\Sigma} - C_{0n}) < 0$ in (13), i.e., $C_{0n}/C_{0\Sigma} < 1 - 1/4\omega^2LC_{0\Sigma}$, $\Delta i_{0n,dist}^{(\varphi)}$ will have the same sign with $\Delta i_{0f,dist}^{(\varphi)}$. As indicated, $\omega^2LC_{0\Sigma} \in [0.9535, 1]$, so the negative superposition can be guaranteed when $C_{0n}/C_{0\Sigma} < 0.738$. It is generally satisfied in today's multi-feeder distribution networks, and also with the fact that HIFs mostly happen in overhead lines whose grounding capacitances are much lower than cables.

In summary, the superposition features of healthy and transformer feeders are completely different from that of faulty feeders. In Fig.5, two field HIFs with different distortion extents validate the above conclusion. The topology and fault position have been introduced in Fig.1.

If distortion offset is considered, $f(\omega t)_{0f}^{(\varphi-\pi)}$ in Fig.4 will not be axial-symmetric. In other words, (7) is not satisfied but (8) still is. Meanwhile, corresponding derivations in (9)-(13) are still valid. Therefore, the above superposition features of different feeders still hold, but the distortions will become non-axial-symmetric like Fig.5(b).

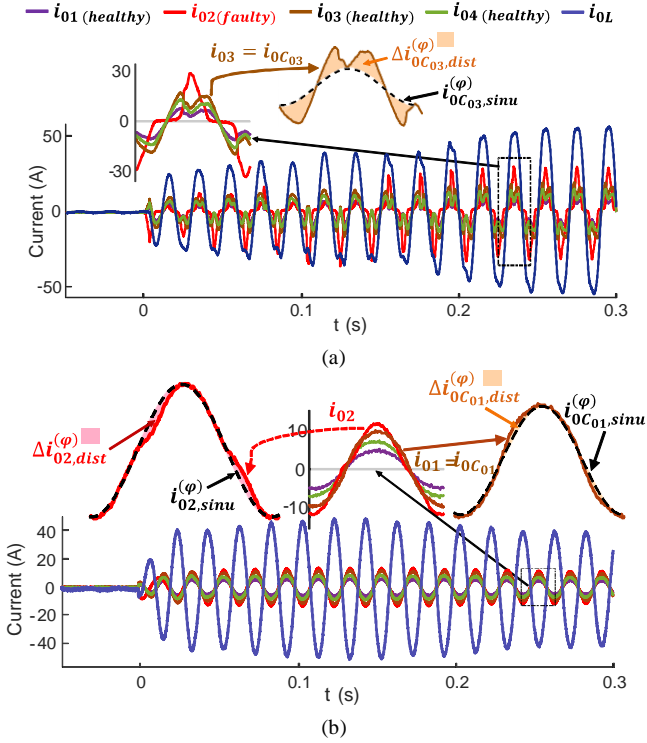

Fig.5 Zero-sequence currents of different feeders when field HIF happens in a 10kV resonant system: HIF grounded to (a) wet soil and (b) wet cement.

### B. Isolated Neutral

When neutral changes, the relationship between currents is different. Let $i_{0L}$ still represent the zero-sequence current of transformer feeder, and it becomes in phase with $i_{0C_{0i}}$ for a network with isolated neutral, which are both set as $\varphi$.

For an isolated neutral network, (9) becomes a linear equation, and the zero-sequence currents of different feeders can be solved as (14). It indicates that the currents of all feeders are obtained by negative superpositions. However, the phases of currents in healthy and faulty feeders are completely opposite.

$$i_{0i\,(i=1,2,\dots,n-1,L)} = i_{0C_{0i},sinu}^{(\varphi)} + \Delta i_{0C_{0i},dist}^{(\varphi)}$$
$$= i_{0C_{0i},sinu}^{(\varphi)} + \frac{C_{0i}}{C_{0\Sigma}} \Delta i_{0f,dist}^{(\varphi)}$$
$$i_{0n} = i_{0n,sinu}^{(\varphi-\pi)} + \Delta i_{0n,dist}^{(\varphi-\pi)} = i_{0n,sinu}^{(\varphi-\pi)} + \left(\frac{\sum_{i=1}^{n-1} C_{0i}}{C_{0\Sigma}}\right) \Delta i_{0f,dist}^{(\varphi-\pi)} \quad (14)$$

### C. Low-Resistor-Earthed Neutral

For this neutral, $i_{0L}$ becomes the current flowing through the equivalent neutral zero-sequence resistor $R_N$ and lags $\pi/2$ behind $i_{0C_{0i}}$. Therefore, the phase of $i_{0f}$ is $\varphi - \theta$ ($0 < \theta < \frac{\pi}{2}$) when that of $i_{0C_{0i}}$ is denoted as $\varphi$. Similar to (6), simplify $\Delta i_{0f,dist}^{(\varphi-\theta)}$ as $f(\omega t)_{0f}^{(\varphi-\theta)}$, and its equation in the range of $\omega t \in [-\varphi + \theta, -\varphi + \theta + \frac{\pi}{2}]$ is expressed as $-e^{\frac{\tau}{\omega}(\omega t+k)} \cdot I_{fM,dist} \sin[2(\omega t + \varphi - \theta)]$. Then, (9) is transformed into a first-order NHL equation:

$$\Delta i_{0L,dist}^{(\varphi-\frac{\pi}{2})} + R_N C_{0\Sigma} \frac{d\Delta i_{0L,dist}^{(\varphi-\frac{\pi}{2})}}{dt} = -e^{\frac{\tau}{\omega}(\omega t+k)} \cdot I_{fM,dist} \sin[2(\omega t + \varphi - \theta)] \quad (15)$$

where, $-1 < \tau < 0$ and $k = \varphi - \theta$. The zero-sequence currents of different feeders are calculated as (16)-(18). Detailed derivations are omitted due to the page limitation.

$$i_{0L} = i_{0L,sinu}^{(\varphi-\frac{\pi}{2})} - I_{fM,dist} \cdot e^{\frac{\tau}{\omega}(\omega t+k)} \sin 2(\omega t + \varphi - \theta)$$
$$= i_{0L,sinu}^{(\varphi-\frac{\pi}{2})} + \Delta i_{0f,dist}^{(\varphi-\theta)} \quad (16)$$

$$i_{0i\,(i\neq n)} = i_{0C_{0i},sinu}^{(\varphi)} - I_{fM,dist} A_{C_{0i}} \cdot e^{\frac{\tau}{\omega}(\omega t+k)} \cos 2(\omega t + \varphi - \theta)$$
$$= i_{0C_{0i},sinu}^{(\varphi)} + A_{C_{0i}} \Delta i_{0f,dist}^{(\varphi-\theta+\frac{\pi}{4})} \quad (17)$$

$$i_{0n} = i_{0n,sinu}^{(\varphi-\theta'-\pi)} + I_{fM,dist} \cdot e^{\frac{\tau}{\omega}(\omega t+k)} \sin 2(\omega t + \varphi - \theta)\}$$
$$= i_{0n,sinu}^{(\varphi-\theta'-\pi)} + \Delta i_{0f,dist}^{(\varphi-\theta-\pi)} \quad (18)$$

where $A_{C_{0i}} = 2\omega R_N C_{0i}$. $\frac{3\pi}{2} - \theta'$ represents the phase of $i_{0n}$ as shown in Fig.6(a), and $\theta < \theta' < \frac{\pi}{2}$.

Similarly, solve the other three intervals in $[-\varphi + \theta, -\varphi + \theta + 2\pi)$. In most cases, $\theta$ is closer to $\pi/2$ than to 0, i.e., $\frac{\pi}{4} < \theta < \frac{\pi}{2}$. One instance is illustrated in Fig.6(b)-(d) to show the superposition features for different feeder currents. The negative and positive superpositions are filled by two colors, which are determined by the 'zero-crossings' of distorted and sinusoidal components, as marked in Fig.6(b)-(c). It is also observed that the simplified $f(\omega t)_{0X(X=L,C_{0i},n)}^{(phase)}$ show similar tendencies to $\Delta i_{0X,dist}^{(phase)}$. Therefore, the superposition features can be qualitatively analyzed just with $f(\omega t)_{0X}^{(phase)}$. We take one cycle as an example:

1) $i_{0L}$. According to (16), the zero-crossings of distorted components lead $\pi/2 - \theta$ ahead that of the sinusoidal components. Therefore, both positive and negative superpositions exist as in Fig.6(b), and the most distorted positions lead a bit ahead of the zero-crossings of its sinusoidal components.

2) $i_{0i\,(i\neq n)}$. According to (17), the zero-crossings of distorted components lag $\theta - \pi/4$ behind that of the sinusoidal components. It causes negative superpositions to happen between two positive ones at each side of the x-axis. Therefore, the most distorted positions happen near the maximal and minimal values of its sinusoidal component. The distortion shape in Fig.6(c) is thereby like that of the healthy feeders at resonant networks (Fig.5(a)), but they are generated by different types of superposition.

3) $i_{0n}$. According to (18), the zero-crossings of distorted components lead $\theta' - \theta$ ahead of sinusoidal components.

The most distorted positions thereby lead a bit ahead of the zero-crossings of its sinusoidal components. In most cases, $\theta' - \theta$ is small and the distorted shapes of $i_{0n}$ and $i_{0f}$ are similar as shown in Fig.6(d).

In summary, the impacts of distorted components on the faulty feeder are apparently different from the healthy feeders, but less different from that on the transformer feeder. However, phases of currents on the transformer and faulty feeders are vastly different from each other, according to Fig.6(a).

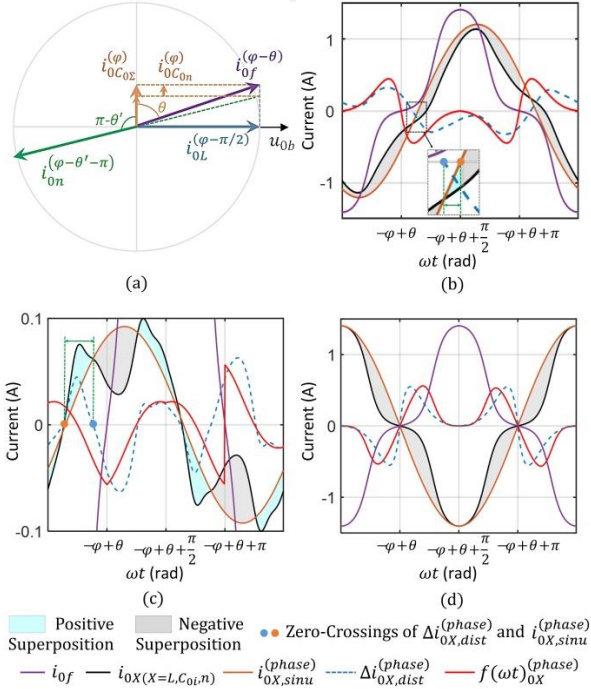

Fig.6 (a) Phase diagram, and relationship between the distorted and sinusoidal components of (b) $i_{0L}$, (c), $i_{0i\,(i\neq n)}$ or $i_{0C_{oi}}$, and (d) $i_{0n}$.

## IV. DETECTION AND FEEDER IDENTIFICATION OF HIF

Based on the distortion features presented by different feeders, this section mainly proposes a feeder identification method for the HIFs in three neutral networks. In the method, distortions of HIFs are described by an interval slope, which is defined in our previous work [29] and used to detect HIFs. To make the paper self-contained, we firstly give a brief introduction to this approach.

The distortion shape of a HIF can be reflected by derivative, however, which is easily interfered with by noises. A definition of interval slope is thereby proposed based on the linear least square fitting (LLSF). For a sampling point $n_s$ of current $i_0(n)$, its interval slope (denoted as $IS_{i_0}(n_s)$) is expressed in (19), representing the slope of a line that linearizes an interval of $i_0(n)$ by LLSF. The interval (denoted as $INT_{n_s}$) lets $n_s$ as the midpoint and with a length of $l$.

$$IS_{i_0}(n_s) = \frac{l \cdot \sum_{INT_{n_s}} [n \cdot i_0(n)] - \sum_{INT_{n_s}} n \cdot \sum_{INT_{n_s}} i_0(n)}{l \cdot \sum_{INT_{n_s}} n^2 - \left(\sum_{INT_{n_s}} n\right)^2} \quad (19)$$

where, $l$ is suggested as $N_T/8$ and $N_T$ represents the number of sampling points in a power frequency cycle.

To further eliminate the impacts of impulse noises or other severe ineffective distortions that could result from intermittent arcs or intense background noises, a robust local regression smoothing (RLRS) combined with the Grubbs Criterion is also proposed in [29]. Then, the values in each interval are refit by the Grubbs-RLRS method before calculating the interval slope. Fig.7 illustrates the processing result of a field HIF, the waveform of which contains the impulse signals generated by intermittent arcs. Comparisons between Fig.7(c) and Fig.7(b) shows the conspicuous advantage of using LLSF over the derivative. It is also presented in Fig.7(a) that the low-pass filter (LPF) with a low cut-off frequency $f_C$ could not eliminate the impulse noise, but enlarge its effects. By combining the Grubbs-RLRS with a higher $f_C$ LPF (around 1500Hz), the elimination of ineffective distortions can be better realized.

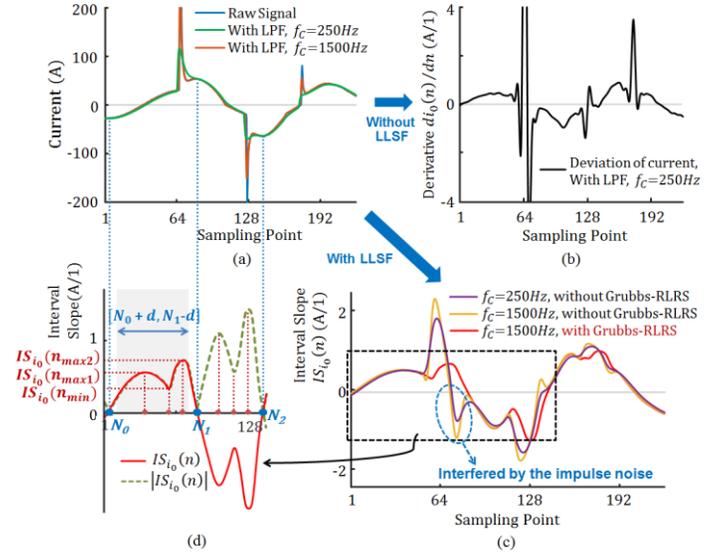

Fig.7 Effectiveness of the LLSF-Grubbs-RLRS method to extract distortions: (a) The zero-sequence current in the faulty feeder of a field HIF in a 10kV system; (b) The derivative of zero-sequence current; (c) The interval slopes with or without Grubbs-RLRS; (d) The 'M shape' for each half-cycle.

With the LLSF-Grubbs-RLRS method, the interval slopes of diverse HIF distortions can be uniformly described as an 'M shape' in each half-cycle, like Fig.7(d). Here we briefly introduce the criterion in [29] to detect HIF distortions. The main idea is to make a judgment about whether the interval slope curve in a half-cycle is exhibited as an 'M shape':

1) In the $[N_0 + d, N_1 - d]$ shown in Fig.7(d), two maximums $|IS_{i_0}(n_{max1})|$, $|IS_{i_0}(n_{max2})|$, and at least one minimum $|IS_{i_0}(n_{min})|$ ($n_{max1} < n_{min} < n_{max1}$) shall exist, where $N_{0,1,2}$ are the zero-crossings of $IS_{i_0}(n)$ obtained by Fourier transform and a certain calibration.

2) All the other minimums that are generated by incomplete filtrations should be between $[n_{max1}, n_{max2}]$ and without large differences of values between each other.

If both two half-cycles satisfy the criteria, the interval slope will present 'double M shape' in a cycle and a 'faulty cycle' is recorded. For any feeder, if there are a few successive cycles recorded as 'faulty cycles', a HIF is detected.

Triggered by the detection of HIF introduced in [29], this paper proposes a method for faulty feeder identification, implemented with synchronous zero-sequence currents of all feeders and the zero-sequence voltage at the substation bus $u_{0b}$. For a half-cycle of $i_{0i}$, an $INDEX_{i_{0i}}$ is defined:

$$INDEX_{i_{0i}} = c_{dir} \cdot \frac{\overline{IS_{i_{0i},nmax}} - IS_{i_{0i}}(n_{min})}{|\overline{IS_{i_{0i},nmax}}|} \quad (20)$$

where, $\overline{IS_{i_{0i},nmax}} = \frac{1}{2}[IS_{i_{0i}}(n_{max1}) + IS_{i_{0i}}(n_{max2})]$, and $c_{dir}$ is a coefficient expressed by the interval slope of $u_{0b}$:

$$c_{dir} = \begin{cases} d\,IS_{u_{0b}}(n_{min})/d\,n & \text{, resonant neutral} \\ -d\,IS_{u_{0b}}(n_{min})/d\,n & \text{, isolated neutral} \\ -IS_{u_{0b}}(n_{min}) & \text{, low resistor earthed neutral} \end{cases} \quad (21)$$

where $c_{dir}$ is further calculated as per-unit values after dividing by its maximum ($c_{dir} \in [0,1]$). Besides, if a cycle is not recorded as 'faulty cycle', $INDEX_{i_{0i}}$ are set as 0 for both half-cycles.

It needs to be claimed that when $i_{0f}$ is extremely small in the resonant neutral network, $i_{0n}$ could be obviously affected by the active components caused by the active losses of Petersen coil and network resistance. Therefore, the phase of $i_{0n}$ would lead that of the healthy feeders by less than $\pi/2$ rad. Under this circumstance, $c_{dir} = -IS_{u_{0b}}(n_{min})$ is more suitable, which means two forms of $c_{dir}$ are both used for the resonant neutral network.

Based on the above descriptions, for all three neutrals, $INDEX_{i_{0i}}$ of faulty feeders is large and positive, whereas $INDEX_{i_{0i}}$ of healthy and transformer feeders are negative or equal to zero.

## V. CASE STUDY

A total of 28 field HIFs, which are grounded to different surface materials and under different humidity in a 10kV network, are used to verify the detection and feeder identification reliability (22 are with pure zero-off distortions and 6 are interfered with by arcing impulse noises). Three instances of HIFs respectively happening in three neutral networks are illustrated in Fig.8. As shown in figure (iv) of Fig.8(a), after 4 successive cycles presenting 'double M shape' and recorded as 'faulty cycles', a HIF is detected. Then, as shown in figure (iii), the feeder identification procedure is triggered and uses data in an 'identification window', which includes 4 cycles before the trigger and more than 20 cycles after that. After choosing $c_{dir}$ according to different neutrals, the $INDEX_{i_{0i}}$ of the faulty feeder shows large and positive values, even for the HIFs with extremely low current amplitudes and severe noise interferences, like the HIF with about 1A current in Fig.8(c).

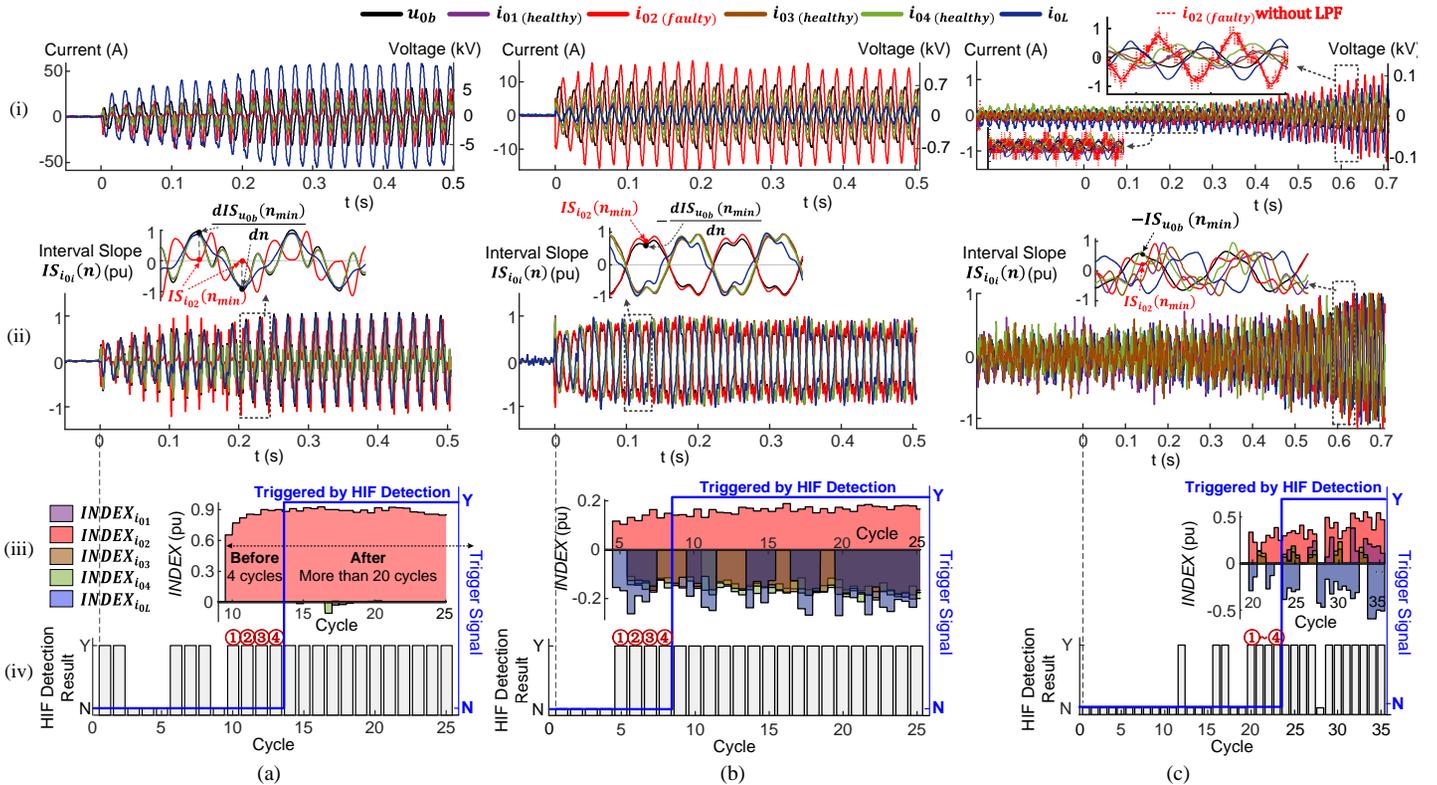

Fig.8 Detection and feeder identification results of the HIFs tested in a 10 kV network with (a) resonant neutral (dry soil), (b) isolated neutral (wet cement), and (c) low-resistor-earthed (10Ω) neutral (dry cement). There are four figures for each result: (i) presents the waveforms of $u_{0b}$, $i_{0i}$ and $i_{0L}$; (ii) exhibits their interval slopes; (iii) shows the $INDEX_{i_{0i}}$ of each feeder, trigger of which is determined by (iv) the detection results of faulty feeder current.

Besides, we quantify the level of $INDEX_{i_{0i}}$ by its mean value within the 'identification window' ($\overline{INDEX_{i_{0i}}}$). After all the 28 HIFs are successively detected, their feeder identification results are shown in Fig.9. In particular, the HIFs of No. 2, 15, 17, 22, 24, and 26 are with current RMS below 6A, while No. 15, 17, and 22 are below 1A. Although indexes are a bit affected, $\overline{INDEX_{i_{0i}}}$ of faulty feeders are still with the largest positive values and HIF feeders can be correctly identified.

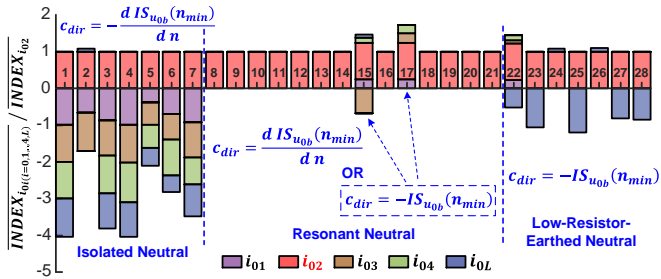

Fig.9 Feeder identification results of 28 HIFs in a 10 kV field network (a stack figure, columns with different colors do not cover each other but are connected separately).

Operations of distributed generations (DGs) would change the directions of power flow during faults, and nonlinear loads could inject harmonics into the system and cause distortions. However, the step-down transformers that connect MV distribution networks and DGs (or loads) are mostly with no-neutral wirings (delta or Y) at the primary side. As a result, the DG penetrations or load behaviors don't affect the zero-sequence currents measured at the side of MV distribution networks [7], [15]. Besides, the capacitor switching and inrush current don't generate zero-off distortions as HIFs, so that the proposed detection and feeder identification algorithms can keep reliability under these non-fault conditions.

## VI. CONCLUSION

Based on the nonlinearities of HIF current, this paper theoretically deduces the differences of distortion features between healthy and faulty feeders, respectively, for distribution networks with three common neutrals. An algorithm for HIF detection and feeder identification is proposed based on the synchronous zero-sequence current at each feeder and the zero-sequence voltage at the substation. The proposed algorithm is feasible for the three network neutrals. In addition, owing to the distortion description and the processing of LLSF-Grubbs-RLRS method, the algorithm is immune to severe noisy environments and able to reliably detect the diverse distortions of HIFs under various fault scenarios.